%% file: main.tex
\documentclass[9pt,journal]{IEEEtran}
\IEEEoverridecommandlockouts
\usepackage[hyphens]{url}
\usepackage{breakurl}
\usepackage{citesort}
\usepackage{amsmath,amsfonts,amsthm,amssymb}
\usepackage{bbm}
\usepackage{algpseudocode}
\usepackage{algorithm}

\usepackage{graphicx}
\usepackage{subfigure}
\graphicspath{{figure/}}
\usepackage{textcomp}
\usepackage{booktabs}
\usepackage{setspace}
\usepackage{tabularx}

\newcolumntype{L}{>{\hspace*{-\tabcolsep}}l}
\newcolumntype{R}{c<{\hspace*{-\tabcolsep}}}
\usepackage[table]{xcolor}
\usepackage{xcolor}
\usepackage{extarrows}
\usepackage{times}
\usepackage[justification = centering]{caption}
\usepackage[font=footnotesize,labelfont=footnotesize]{caption}

\usepackage{tikz}
\usetikzlibrary{arrows.meta, positioning, shapes, calc, fit}
\definecolor{lightblue}{rgb}{0.93,0.95,1.0}

\def\BibTeX{{\rm B\kern-.05em{\sc i\kern-.025em b}\kern-.08em
		T\kern-.1667em\lower.7ex\hbox{E}\kern-.125emX}}
\input{newcommand.tex}

\allowdisplaybreaks
\begin{document} 
\captionsetup{justification=raggedright,singlelinecheck=false}
\title{
 Holographic MIMO-assisted Multiuser Transmission with Electromagnetic Exposure Constraints
}
\author{ Yilin Yang, Mengyu Qian,~\IEEEmembership{Graduate Student Member,~IEEE},  Ahmed Elzanaty,~\IEEEmembership{Senior Member,~IEEE},\\ and Li~You,~\IEEEmembership{Senior Member,~IEEE}
\vspace{-0.7cm}
\thanks
{
Copyright (c) 2026 IEEE. Personal use of this material is permitted. However, permission to use this material for any other purposes must be obtained from the IEEE by sending a request to pubs-permissions@ieee.org.

This work was supported by the National Natural Science Foundation of China for Outstanding Young Scholars under Grant 62322104, the Jiangsu Province Major Science and Technology Project under Grant BG2024005, the Natural Science Foundation of Jiangsu Province under Grant BK20231415, and the Fundamental Research Funds for the Central Universities under Grant 2242022k60007. \textit{(Yilin Yang and Mengyu Qian contributed equally to this work.)} \textit{(Corresponding author: Li You.)}

Yilin Yang, Mengyu Qian, and Li~You are with the National Mobile Communications Research Laboratory, Southeast University, Nanjing 210096, China, and also with the Purple Mountain Laboratories, Nanjing 211100, China (e-mail: yilinyang@seu.edu.cn; qianmy@seu.edu.cn; lyou@seu.edu.cn).

Ahmed Elzanaty is with 5GIC \& 6GIC, Institute for Communication Systems (ICS), University of Surrey, GU2 7XH Guildford, U.K. (e-mail:
a.elzanaty@surrey.ac.uk).
}
}

\maketitle

\begin{abstract}
Holographic multiple-input multiple-output (HMIMO) has emerged as a promising technology for future wireless systems by enabling continuous electromagnetic (EM) field control over large apertures. However, user-side EM exposure has become an increasingly important concern in large-scale array systems. This paper addresses this issue by developing a multiuser uplink HMIMO model, where a physically consistent specific absorption rate (SAR) model is established to quantify the EM exposure. On this basis, a spectral efficiency (SE) maximization problem is addressed by developing a modified iterative water-filling algorithm. Simulation results demonstrate that the proposed algorithm effectively improves the system SE while satisfying the SAR constraints.

\end{abstract}

\begin{IEEEkeywords}
	Holographic multiple-input multiple-output (HMIMO), specific absorption rate (SAR), spectral efficiency (SE).
\end{IEEEkeywords}

\section{Introduction}
The sixth-generation (6G) wireless networks are projected to handle the explosive rise in data traffic, driven by a surge of new devices and applications, with far greater spectral efficiency (SE) requirements \cite{huang2020holographic,ye2023fluid}. With the advancement of metamaterials technology, the concept of holographic multiple-input multiple-output (HMIMO) has been explored to enhance the transmission performance with limited antenna aperture \cite{10505154}. Unlike traditional MIMO, HMIMO employs densely distributed sub-wavelength antenna elements that form a quasi-continuous aperture \cite{an2023tutorial}. By precisely shaping the spatial electromagnetic (EM) field, HMIMO can efficiently encode information onto the wavefronts and approach the theoretical capacity bound of a finite aperture \cite{dardari2021holographic}. 

Recent studies have extensively investigated HMIMO in terms of channel modeling, beamforming design and performance analysis. For example, \cite{9906802} proposed a wavenumber division multiplexing scheme based on Fourier basis expansion for one-dimensional transceiver models under line of sight (LoS) conditions. Later, \cite{zhang2023pattern} and \cite{qian2024spectral} further studied multiuser HMIMO transmission and focused on SE optimization for HMIMO systems.

Although these works have demonstralicitly account for EM exposure. This issue is particularly important for HMIMO, since its quasi-continuous aperture can generate highly focused EM fields, which may lead to elevated local energy absorption for users located close to the transmitting surface.
ted the potential performance gains of HMIMO, they mainly considered system designs under transmit power constraints and did not exp

To quantify such exposure, the specific absorption rate (SAR) has been widely adopted as a standard metric, measuring the rate at which EM energy is absorbed by biological tissues \cite{chiaraviglio2021health}. In conventional MIMO systems, SAR-aware transmission design has been extensively studied. Prior works have investigated precoding optimization under SAR constraints \cite{ying2015closed}, \cite{9730856}. More recently, these designs have been extended to emerging architectures such as hybrid dynamic metasurface antenna (DMA)-assisted systems \cite{jiang2022hybrid} and fluid antenna system (FAS)-assisted MIMO \cite{11080242}, where SAR constraints are incorporated into joint system optimization frameworks.

Nevertheless, the SAR-aware design methodologies developed for conventional systems are fundamentally based on discrete antenna models, where transmit signals are represented by finite-dimensional vectors and SAR is characterized through predefined quadratic forms. In contrast, HMIMO systems are governed by continuous EM field propagation over quasi-continuous apertures, where both signal representation and EM exposure are inherently spatially continuous. These fundamental differences prevent the direct application of existing SAR-constrained MIMO frameworks and call for new modeling and design approaches.

Inspired by current works, the main contributions of this paper are summarized as follows:
\begin{itemize}
\item{Unlike conventional SAR models based on predefined matrices, we develop a Green’s-function-based EM channel model and, for the first time, derive a physically consistent SAR formulation that explicitly captures the quasi-continuous current distribution over holographic apertures.}
\item{ We further show that the SAR constraints can be expressed as quadratic forms of expansion coefficients and equivalently reformulated as linear trace constraints with respect to the transmit covariance matrix, thereby preserving convexity.}
\item{We propose a modified iterative water-filling algorithm that jointly accounts for transmit power and SAR constraints. By introducing a whitening transformation and generalized eigenvalue decomposition, the proposed method enables efficient transmit covariance optimization beyond conventional power backoff strategies.}
\end{itemize}

\textit{Notations}: Scalars, vectors/matrices, and Euclidean subspaces are represented by regular, boldface, and calligraphic letters, respectively. The sets of integer, complex and real numbers are denoted by $\mathbb{Z}$, $\mathbb{C}$, $\mathbb{R}$, respectively. The inverse, conjugate, transpose, conjugate transpose, and trace operator are denoted by $(\cdot)^{-1}$, $(\cdot)^*$, $(\cdot)^T$, $(\cdot)^H$, and $\mathrm{tr}(\cdot)$, respectively. The diagonalization of matrix $\mathbf{A}$ is denoted by $\mathrm{diag}(\mathbf{A})$, which constructs a diagonal matrix from vectors. The Lebesgue measure of $\mathcal A$ is $|\mathcal A|$. The Euclidean norm is denoted by $\|\cdot\|$. The expectation operator is denoted by $\mathbb{E}\{\cdot\}$. The Kronecker product is denoted by $\otimes$. The imaginary unit is denoted by $\jmath =\sqrt{-1}$. The notation $\lceil x \rceil$ denotes taking the smallest integer greater than or equal to $x$, and $(a)^+=\max\{a,0\}$.

\section{System Model}\label{sec:System Model}

\subsection{HMIMO-assisted Model}
We consider a $K$-user uplink HMIMO system, where the base station (BS) is equipped with a holographic planar array (HPA) $\Gamma$ with size $L_{r,x} \times L_{r,y}$ situated on the $x$–$y$ plane with its center at the origin. Each user $k$ is equipped with a transmitting HPA $\Omega_k$, where $k \in \mathcal K$ and $\mathcal K$ represents the set of users. Since different users may have arbitrary orientations, their HPAs may not align with the BS reference plane. To establish a consistent spatial description, the geometry of each transmitting HPA is expressed in the global coordinate system through a rotation characterized by an orthogonal matrix $\mathbf{R}_k\in\!\mathbb{R}^{3\times3}$. Let ${\mathbf{u}_k}:=({u}_{k,x},{u}_{k,y},{u}_{k,z})\in\!\Omega_k$ denote a point on the local surface of the $k$-th transmitting HPA. Its corresponding coordinate in the global coordinate system is written as $\hat{\mathbf{u}}_k = \mathbf{R}_k{\mathbf{u}_k}$, and $\hat{\Omega}_k \triangleq \{\mathbf{R}_k{\mathbf{u}_k}\!:\!\mathbf{u}_k\! \in \!\Omega_k \}$ represents the set of coordinates of plane $\Omega_k$ in the global coordinate system. The projections of $\hat{\Omega}_k$ along the global $x$ and $y$ axes are $L_{k,x}$ and $L_{k,y}$.

Although it is beneficial to exploit multiple polarization states, it will also lead to higher hardware design complexity. Hence, in this paper, the transmitter is assumed to be vertically polarized, such that the source current excitation is limited to its $y$-component. Assuming that the $k$-th user transmits $L_k$ data streams, then the transmit signal can be written as
\begin{align}\label{2.2}
	\mathbf{X}_k({\mathbf{u}}_k) =\sum_{\ell = 1}^{L_k} {X}_{k,\ell}({\mathbf{u}}_k) {\mathbf{e}}_{k,y}  s_{k,\ell},
\end{align}
where ${X}_{k,\ell}({\mathbf{u}}_k)$ denotes the scalar current density allocated to the $\ell$-th stream in the local coordinates and the polarization vector in the local coordinates is ${\mathbf{e}}_{k,y}=[0,1,0]^T$. Moreover, $s_{k,\ell}$ denotes the signal carried by the current density ${X}_{k,\ell}({\mathbf{u}}_k)$. For simplicity, it is assumed that $s_{k,\ell}$ satisfies $\mathbb{E}\{|s_{k,\ell}|^2\} = 1$, $\mathbb{E}\{s_{k,\ell} s^*_{k,\ell'}\} = 0 , \forall \ell' \neq \ell$, $\mathbb{E}\{s_{k,\ell}s^* _{k',\ell}\} = 0, \forall k' \neq k$.

According to \cite{chew1995waves}, the EM field induced at a receiving point $\mathbf{v}$ by the current density $\mathbf{X}_k({\mathbf{u}}_k)$ follows Maxwell’s equations. Hence, the electric field received at a point $\mathbf{v}\!\in\!\Gamma$ on the BS aperture due to user $k$ is expressed as
\begin{equation}
\mathbf{E}_k(\mathbf{v}) =
\int_{\hat{\Omega}_k}
\mathbf{G}(\mathbf{v},{ \hat{\mathbf{u}}_k})
\mathbf{X}_{k}(\hat{\mathbf{u}}_k)\,\mathrm d \hat{\mathbf{u}}_k ,
\label{eq:field_integral}
\end{equation}
where $\mathbf{G}(\mathbf{v},{ \hat{\mathbf{u}}_k})$ is the Green’s function for free space propagation. It serves as an idealized EM propagation model and is adopted to enable a tractable and physically consistent system representation. Since we only consider the radiation field, the higher order terms of the Green’s function can be omitted \cite{franceschetti2017wave}, i.e.,
\begin{align}\label{Green}
	\mathbf{G}(\mathbf{v},{ \hat{\mathbf{u}}_k}) \approx &-\frac{\jmath \eta \exp \left(-\jmath k_0 \|\mathbf{v}-{ \hat{\mathbf{u}}_k}\|\right)}{2 \lambda \|\mathbf{v}-{ \hat{\mathbf{u}}_k}\|}\left(\mathbf{I}-\hat{\mathbf{p}} \hat{\mathbf{p}}^H\right),
\end{align}
where $k_0 = 2\pi/\lambda$ represents the free space wavenumber; $\lambda = c/f$ is the wavelength of the EM wave in free space, with $c$ and $f$ denoting the speed of light and the carrier frequency, respectively; $\eta$ signifies the wave impedance, and $\hat{\mathbf{p}} := (\mathbf{v} - \hat{\mathbf{u}}_k) / ||\mathbf{v} - \hat{\mathbf{u}}_k||$.
Additionally, $\mathbf{X}_k(\hat{\mathbf{u}}_k)$ denotes the representation of the transmit signal in the global coordinate system, which is expressed as
\begin{equation}\label{eq:xk}
\mathbf{X}_k(\hat{\mathbf u}_k)
=\sum_{\ell=1}^{L_k} X_{k,\ell}(\mathbf{R}_k^T\hat{\mathbf{u}}_k)\, \mathbf{R}_k^{T}\hat{\mathbf e}_{k,y} s_{k,\ell},
\end{equation}
where $\hat{\mathbf e}_{k,y} = \mathbf{R}_k{\mathbf e}_{k,y}$ denotes the global-coordinate representation of the local polarization unit vector.

Building on the previous discussion, the received field at the BS aperture is then given by
\begin{align}
{Y}(\mathbf{v}) &=
\sum_{k=1}^{K} \hat{\mathbf{e}}_{k,y}^T \mathbf{E}_k(\mathbf{v})+ {n}(\mathbf{v})\\\notag
&= \sum_{k=1}^{K} \hat{\mathbf{e}}_{k,y}^T \int_{\hat{\Omega}_k}\mathbf{G}(\mathbf{v}, \hat{ \mathbf{u}}_k)\mathbf{X}_{k}(\hat{ \mathbf{u}}_k)\,\mathrm d \hat{ \mathbf{u}}_k+ {n}(\mathbf{v})\\\notag
&= \sum_{k=1}^{K}\sum_{\ell=1}^{L_k}\int_{\hat{\Omega}_k}h(\mathbf{v},\hat{\mathbf{u}}_k)X_{k,\ell}(\mathbf{R}_k^T\hat{\mathbf{u}}_k)s_{k,\ell}\mathrm d{ \hat{\mathbf{u}}_k}+ {n}(\mathbf{v}),
\label{eq:Yv}
\end{align}
where $h(\mathbf{v},\hat{\mathbf{u}}_k)\! = \!\hat{\mathbf{e}}_{k,y}^T\mathbf{G}(\mathbf{v},\hat{ \mathbf{u}}_k)\mathbf{R}_k^T\hat{\mathbf{e}}_{k,y}\! \in \!\mathbb{C}$, and $ n(\mathbf v)$ denotes the additive receiver noise field over $\Gamma$, modeled as a zero-mean circularly symmetric complex Gaussian random field.
\vspace{-10pt}

\subsection{Fourier Series Expansion}
Exact eigenfunction decomposition of the spatial channel is computationally prohibitive for large HMIMO systems \cite{miller2000communicating}. We therefore employ Fourier expansion, which is suitable for representing the induced current using compact basis functions, while keeping its intrinsic features. 

For user $k$, the discrete spatial frequencies are given by
\begin{align}
\kappa_{k,nx} = \frac{2\pi n_x}{L_{k,x}}, 
\qquad 
\kappa_{k,ny} = \frac{2\pi n_y}{L_{k,y}}, 
\quad n_x,n_y \in \mathbb{Z},
\end{align}
where $n_x$ and $n_y$ are the integer indices in the $x$ and $y$ dimensions, respectively. Since $\kappa_{k,n_x}^2 + \kappa_{k,n_y}^2 \leq \kappa_0^2$ \cite{ji2023extra}, the index set is finite, which is given by
\begin{align}
\mathcal{N}_k=\left\{\!(n_x,n_y)\in\mathbb{Z}^2\!:\!|n_x|\!\leq\!\left\lceil\frac{L_{k,x}}{\lambda}\right\rceil\!,\!|n_y|\!\leq\!\left\lceil\frac{L_{k,y}}{\lambda}\right\rceil\right\}.
\end{align}

Based on this, the current distribution for user $k$ is given by
\begin{align}\label{xkl}
	{X}_{k,\ell}(\mathbf{R}_k^T\hat{\mathbf{u}}_k) &= \sum_{\mathbf{n} \in \mathcal{N}_k} {\xi}_{k,\ell,\mathbf{n}}\phi_{k,\mathbf{n}}(\mathbf{R}_k^T\hat{\mathbf{u}}_k),
\end{align}
where $\mathbf{n}:=(n_x,n_y)$, ${\xi}_{k,\ell,\mathbf{n}}$ is the expansion coefficient, which corresponds to the projection weights of the transmit current distribution onto the orthogonal Fourier basis functions, and
\begin{align}
	\phi_{k,\mathbf{n}}(\mathbf{R}_k^T\hat{\mathbf{u}}_k)\!= \!\frac{e^{\jmath \left(\kappa_{k,n_x}\hat{\mathbf{e}}_{k,x}^T\mathbf{R}_k^T\hat{\mathbf{u}}_{k}\! +\!\kappa_{k,n_y}\hat{\mathbf{e}}_{k,y}^T\mathbf{R}_k^T\hat{\mathbf{u}}_{k}\right)}}{\sqrt{L_{k,x} L_{k,y}}}
\end{align}
denotes the 2D Fourier basis function. The dimension of the input space for the $k$-th user is expressed as $N_k = |\mathcal{N}_k| = \left( 2\frac{L_{k,x}}{\lambda} + 1 \right)\left( 2\frac{L_{k,y}}{\lambda} + 1 \right)$.

Similarly, the received field over $\Gamma$ is projected onto the wavenumber domain
by employing a set of Fourier basis functions $\{\psi_{\mathbf m}(\mathbf v)\}$,
\begin{equation}
\psi_{\mathbf m}(\mathbf v)
=
\frac{1}{\sqrt{L_{r,x}L_{r,y}}}
e^{\jmath\left(
\frac{2\pi}{L_{r,x}} m_x v_x
+
\frac{2\pi}{L_{r,y}} m_y v_y
\right)}.
\end{equation}
Here, $\mathbf m=(m_x,m_y)\in\mathcal M$, with $m_x$ and $m_y$ corresponding to the integer indices in the $x$ and $y$ dimensions. The dimension of the output space is given by $M = |\mathcal{M}| = \left( 2\frac{L_{r,x}}{\lambda} + 1 \right)\left( 2\frac{L_{r,y}}{\lambda} + 1 \right)$ .

Thus, the input-output relationship is reduced to a finite-dimensional form that can be efficiently analyzed:
\begin{align}
	{y}_j = \sum_{k=1}^{K}\sum_{\ell=1}^{L_k}\sum_{i=1}^{N_k} {H}_{k,j,i}{\xi}_{k,\ell,i}  s_{k,\ell} + {n}_j, j \in M,
\end{align}
where the effective channel coefficients are
\begin{align}\label{H_kji}
	{H}_{k,j,i} = \int_{\Gamma} \int_{\hat{\Omega}_k} \psi^{*}_{\mathbf{m}_j}(\mathbf{v})h(\mathbf{v},\hat{\mathbf{u}}_k)\phi_{k,\mathbf{n}_i}(\mathbf{R}_k^T\hat{\mathbf{u}}_k) {\mathrm{d}\hat{\mathbf{u}}_k} {\mathrm{d}\mathbf{v}},
\end{align}
and the projection of noise ${n}(\mathbf{v})$ along the $j$-th receiving spatial basis function $\psi_{\mathbf{m}_j}(\mathbf{v})$ is $n_j$,
\begin{align}
n_j
=
\int_{\Gamma}
\psi_{\mathbf m_j}^*(\mathbf v)\, n(\mathbf v)\,\mathrm d\mathbf v , 
n_j \sim \mathcal{CN}(0,\sigma^2).
\end{align}
Therefore, by optimizing the Fourier expansion coefficients $\{\xi_{k,\ell,i}\}$ in the above finite-dimensional representation, the desired current distribution can be directly synthesized.

\subsection{EM Exposure Model}\label{sec:Electromagnetic Exposure Model}
In practical uplink wireless systems, the achievable SE is jointly affected by the transmit power budget and the imposed SAR limits through their impact on the induced current density. In HMIMO scenario, the power-related constraints can be formulated as \cite{qian2024spectral}
\begin{align}\label{transmit power_constraint}
	\int_{\hat{\Omega}_k} ||\mathbf{X}_k(\hat{\mathbf{u}}_k)||^2 {\mathrm{d}\hat{\mathbf{u}}_k} =  \sum_{\ell=1}^{L_k} \int_{\hat{\Omega}_k} &|{X}_{k,\ell}(\mathbf{R}_k^T\hat{\mathbf{u}}_k)|^2  {\mathrm{d}\hat{\mathbf{u}}_k}\notag\\ &\leq P_{k,\rm max},\forall k \in \mathcal{K},
\end{align}
where $P_{k,\rm max}$ represents the power budget allocated to the $k$-th user. Using the Fourier basis function expansion, the above power constraint can be further expressed as
\begin{align}\label{power constraint}
	\sum_{\ell=1}^{L_k}\sum_{i=1}^{N_k} |{\xi}_{k,\ell,i} s_{k,\ell}|^2 
	\!=\!\sum_{\ell=1}^{L_k}\!\sum_{i=1}^{N_k}\! |{\xi}_{k,\ell,i}|^2 \!\leq \!P_{k,\max},\forall k \in \mathcal{K}.
\end{align}

Furthermore, SAR is typically used to quantify the EM exposure experienced by users, representing the rate at which EM energy is absorbed by biological tissues per unit mass. This metric is the most commonly employed measure for evaluating EM exposure. The definition of SAR is given by
\begin{align}
	\mathrm{SAR}(\mathbf{r})=\frac{\chi_{\mathrm t}|\mathbf{E}_m(\mathbf{r})|^2}{2\rho}.
\end{align}
Here, $\chi_{\mathrm t}$ represents the tissue conductivity, $\rho$ denotes the mass density, and $\mathbf{E}_m(\mathbf{r})$ refers to the electric field strength at the position $\mathbf{r}$ within the testing volume \cite{lin2002specific}.

Unlike traditional SAR models that use a predefined SAR matrix, the SAR model in HMIMO system accounts for continuous surfaces and complex mutual coupling, making a predefined matrix impractical. Specifically, for a given test volume $V$, where both $V$ and $\mathbf{r} \in V$ are defined in the global coordinate, based on (\ref{eq:field_integral}), the spatial average value of SAR can be expressed as
\begin{subequations}
\begin{align} 
	\mathrm{SAR}_{k,V}  & = \frac{\chi_{\mathrm t}}{2\rho|V|} \int_{V} \left\| \mathbf{E}_{k}(\mathbf{r}) \right\|^{2} \, \mathrm{d} \mathbf{r} \label{SARkV}\\ 
	\label{eq:SAR}
         & = \frac{\chi_{\mathrm t}}{2\rho|V|} \int_{V} \left( \int_{\hat{\Omega}_k} \mathbf{G}(\mathbf r , \hat{\mathbf u}_k) \mathbf{X}_k(\hat{\mathbf u}_k) \, \mathrm{d}\hat{\mathbf u}_k \right)^H \\ \notag 
         & \quad \left( \int_{\hat{\Omega}_k} \mathbf{G}(\mathbf r , \hat{\mathbf u}_k) \mathbf{X}_k(\hat{\mathbf u}_k) \, \mathrm{d}\hat{\mathbf u}_k \right) \, \mathrm{d}\mathbf{r}.
\end{align}
\end{subequations}

Substituting (\ref{eq:xk}) and (\ref{xkl}) into (\ref{eq:SAR}), and leveraging the linearity of EM propagation under Maxwell’s equations, the electric field at location $\mathbf r \in V$ can be expressed as a linear superposition of mode contributions:
\begin{equation}\label{ekr}
\mathbf E_k(\mathbf r) =  \sum_{\ell=1}^{L_k}\sum_{i=1}^{N_k} \xi_{k,\ell,i}\, \mathbf w_{k,i}(\mathbf r)s_{k,\ell},
\end{equation}
where the contribution of the $i$-th transmit mode is defined as
\begin{equation}\label{wki}
\mathbf w_{k,i}(\mathbf r) \triangleq \int_{\hat{\Omega}_k} \mathbf G(\mathbf r,\hat{\mathbf u}_k)\mathbf R_k^T \hat{\mathbf e}_y\, \phi_{k,n_i}(\mathbf R_k^T \hat{\mathbf u}_k)\, d\hat{\mathbf u}_k.
\end{equation}
Substituting (\ref{ekr}) into the SAR expression in (\ref{SARkV}), and noting that SAR depends on the squared magnitude of the electric field, the SAR over the volume $V$ becomes a quadratic form with respect to the expansion coefficients:
\begin{equation}
\mathrm{SAR}_{k,V} = \sum_{\ell=1}^{L_k}\sum_{i,i'} \xi_{k,\ell,i}^* S_{k,V}^{(i,i')} \xi_{k,\ell,i'},
\end{equation}
where the $(i,i')$-th element of SAR coupling matrix $\mathbf S_{k,V} \in \mathbb{C}^{N_k \times N_k}$ is defined as
\begin{equation}\label{sv}
S_{k,V}^{(i,i')} \triangleq \frac{\chi_{\mathrm t}}{2\rho |V|} \int_V \mathbf w_{k,i}^H(\mathbf r)\mathbf w_{k,i'}(\mathbf r)\, d\mathbf r.
\end{equation}
By introducing the transmit covariance matrix $\boldsymbol{\Xi}_{k,\ell} = \boldsymbol{\xi}_{k,\ell}\boldsymbol{\xi}_{k,\ell}^H$, the SAR expression can be  rewritten in a linear trace form as
\begin{equation}\label{SAR}
\mathrm{SAR}_{k,V} = \sum_{\ell=1}^{L_k} \mathrm{tr}\!\left(\boldsymbol{\Xi}_{k,\ell} \mathbf S_{k,V}\right).
\end{equation}
This reformulation relies on the linearity of the EM field and the quadratic definition of SAR, which together ensure that the SAR constraint is affine in the transmit covariance matrix. According to (\ref{SAR}), the SAR value is determined by the distribution of induced current and the coupling matrix. As the transmit power increases, the induced current density also increases, which will lead to higher SAR value. However, excessive power can exceed the safety limit of SAR, thereby affecting the performance of the system. Therefore, the system needs to find a balance between power and SAR constraints.

Moreover, the analytical expression of the SAR coupling matrix $\mathbf{S}_{k,V}$ in (\ref{sv}) involves a three-dimensional volume integral over the test region $V$, which is generally difficult to handle. To evaluate $\mathbf{S}_{k,V}$ numerically while preserving physical accuracy, we employ a volume discretization approach.

Specifically, the test volume $V$ can be uniformly partitioned into $Q$ small voxels. Let the center of the $q$-th voxel be denoted as $\mathbf{v}_q$ and its volume as $\Delta V_q$. Then, the continuous integral in (\ref{sv}) can be approximated by a Riemann sum as
\begin{align}
{S}_{k,V}^{(i,i')} \approx 
\frac{\sigma}{2\rho |V|}
\sum_{q=1}^Q 
\mathbf w_{k,i}^H(\mathbf{v}_q)\, 
\mathbf w_{k,i'}(\mathbf{v}_q)\, 
\Delta V_q,
\label{eq:SV_sum}
\end{align}
where $\mathbf w_{k,i}(\mathbf{v}_q) = \int_{\hat{\Omega}_k} \mathbf G (\mathbf{v}_q,\hat{\mathbf u}_k)\mathbf{R}_k^{T}\hat{\mathbf e}_y \phi_{k,\mathbf{n}_i}(\hat{\mathbf u}_k)\, \mathrm d \hat{\mathbf u}_k$ denotes the field response at voxel $\mathbf{v}_q$ excited by the $i$-th Fourier mode. By defining $\tilde{\mathbf w}_{k,i} \triangleq \big[ {\mathbf{w}_{k,i}^T(\mathbf{v}_1}),\, {\mathbf{w}_{k,i}^T(\mathbf{v}_2}),\, \ldots,\, {\mathbf{w}_{k,i}^T(\mathbf{v}_Q}) \big]^T $, and collecting them into the matrix $\mathbf{F}
= \big[ \tilde{\mathbf w}_{k,1},\, \tilde{\mathbf w}_{k,2},\, \ldots,\, \tilde{\mathbf w}_{k,N_k} \big]
\in\mathbb C^{3Q\times N_k}$, we then introduce a diagonal weight matrix $\mathbf{W}= \mathrm{diag}(\Delta V_1, \Delta V_2 , \ldots, \Delta V_Q)\otimes\mathbf I_3
\in\mathbb R^{3Q\times 3Q}$ to incorporate the volume contribution of each voxel. With this notation, the discretized SAR matrix can be written in a compact quadratic form:
\begin{align}
\mathbf{S}_{k,V} \approx 
\frac{\sigma}{2\rho |V|}\, 
\mathbf{F}^{H} \mathbf{W} \mathbf{F}.
\label{eq:SV_matrix}
\end{align}

This formulation transforms the continuous volume integral into a weighted matrix product that is straightforward to compute numerically. Moreover, the resulting matrix $\mathbf{S}_{k,V}$ is Hermitian and positive semidefinite by construction, which preserves the physical interpretation of SAR as a non-negative absorbed power density.

\section{Problem Formulation and EM Aware SE Optimization}
In this section, we formulate the SE maximization problem subject to power and SAR constraints, with the optimization variables being the expansion coefficients of each transmission stream. A structured iterative water-filling algorithm is then derived to solve this problem.
\subsection{Problem Formulation}
Based on the basis representation of the induced current introduced in Section II, the achievable SE can be expressed as
\begin{align}
  R_\mathrm{sum} = \mathrm{log}_2(1 + \sum_{k=1}^K{|{\alpha}_k|^2}/\sigma^2),
\end{align}
where ${\alpha}_k$ can be expressed as
\begin{align}
  {\alpha}_k = \sum_{\ell=1}^{L_k} \int_{\hat{\Omega}_k} h(\mathbf{v},\hat{\mathbf{u}}_k)X_{k,\ell}(\mathbf{R}_k^T\hat{\mathbf{u}}_k) \mathrm{d}\hat{\mathbf{u}}_k,
\end{align}
which represents the corresponding amplitude of the $k$-th user due to the multiple induced data streams.

Define the channel matrix $\mathbf{H}_k \!\in\! \mathbb{C}^{M \times N_k}$, where the $(j,i)$-th element is $H_{k,j,i} \in \mathbb{C}$, as specified in (\ref{H_kji}). Furthermore, introduce $\boldsymbol{\Xi}_k = \mathrm{diag}\{\boldsymbol{\Xi}_{k,1}, \boldsymbol{\Xi}_{k,2}, \dots, \boldsymbol{\Xi}_{k,L_k}\}$, and let $\hat{\mathbf{H}}_k = [\mathbf{H}_k, \mathbf{H}_k, \dots, \mathbf{H}_k]$. Additionally, define $\widetilde{\mathbf{S}}_{k,V} \triangleq \mathbf{I}_{L_k} \otimes \mathbf{S}_{k,V}$. Finally, let $Q_{\mathrm{SAR},k}$ denote the SAR budget of the system. The optimization problem is thus defined as
\begin{subequations}\label{sum rate A}
	\begin{align}
		\mathcal{P}_A: \quad \max_{\boldsymbol{\Xi_k}}&  \quad 
\log_2\det\!\left(
\mathbf{I}_M + \frac{1}{\sigma^2} \sum_{k=1}^K \hat{\mathbf{H}}_k\boldsymbol{\Xi}_k\hat{\mathbf{H}}_k^H\right), \\
		\mathrm{s.t.}&  \quad \mathrm{tr}({\boldsymbol{\Xi}}_k) \leq P_{k,\max},  \forall k \in \mathcal{K},\\
		&  \quad \mathrm{tr}({{\boldsymbol{\Xi}}_{k}}\widetilde{\mathbf S}_{k,V})  \leq Q_{\mathrm{SAR},k},  \forall k \in \mathcal{K}.
	\end{align}
\end{subequations}

\subsection{EM Aware SE Optimization}\label{sec:Optimization}
Based on the previous system model, we formulate the problem of maximizing the system SE in $\mathcal{P}_A$.

\subsubsection{Lagrangian Construction and Dual Decomposition}
To incorporate these constraints, we introduce non-negative dual variables $\lambda_k \ge 0$ and $\mu_k \ge 0$. Then, the Lagrangian is
\begin{align}
\mathcal{L}&(\{\boldsymbol{\Xi}_k\}_{k \in \mathcal K},\{\lambda_k,\mu_k\}_{k \in \mathcal K})\nonumber\\
= &
\log_2\det\!\left(
   \! \sigma^2\mathbf{I}_M 
    \!+\! \sum_{k=1}^K \hat{\mathbf{H}}_k\boldsymbol{\Xi}_k\hat{\mathbf{H}}_k^{H}
\!\right) 
\!- \!\sum_{k=1}^K 
  \! \lambda_k\!\left(
      \mathrm{tr}(\boldsymbol{\Xi}_k)\! -\!P_{k,\max}\right) \nonumber\\
-& \sum_{k=1}^K 
   \mu_k\!\left(
      \mathrm{tr}(\widetilde{\mathbf{S}}_{k,V}\boldsymbol{\Xi}_k)\!
      - Q_{\mathrm{SAR},k}
   \right).
\end{align}

It can be observed that both the transmit power and SAR constraints are affine functions of $\boldsymbol{\Xi}_k$. As a result, the dual variables $\lambda_k$ and $\mu_k$ appear in the Lagrangian only through the linear combination $\mathbf A_k = \lambda_k \mathbf I + \mu_k \widetilde{\mathbf S}_{k,V}$ and $ C_k = \lambda_k P_{k,\max} + \mu_k Q_{\mathrm{SAR},k}$. Using these definitions, the Lagrangian can be rewritten as
\begin{align}
\mathcal{L}=\log_2\det (\sigma^2 \mathbf{I}_M & + \sum_{k=1}^K  \hat{\mathbf{H}}_k \boldsymbol{\Xi}_k \hat{\mathbf{H}}_k^{H}) \nonumber\\
&- \sum_{k=1}^K \mathrm{tr}(\mathbf{A}_k \boldsymbol{\Xi}_k) + \sum_{k=1}^K C_k .
\end{align}

The dual function is
\begin{align}\label{dualfunction}
g(\{\lambda_k,\mu_k\}_{k \in \mathcal K})
=
\max_{\boldsymbol{\Xi}_k\succeq0}
\mathcal{L}(\{\boldsymbol{\Xi}_k\}_{k \in \mathcal K},\{\lambda_k,\mu_k\}_{k \in \mathcal K}),
\end{align}
and the dual problem can be written as
\begin{align}\label{dualproblem}
\min_{\lambda_k\ge 0,\,\mu_k\ge 0}~ g(\{\lambda_k,\mu_k\}_{k \in \mathcal K}).
\end{align}

\subsubsection{Transmit Covariance Optimization via Whitening and Water-Filling}
With the Lagrangian and dual function already established in the previous section, the inner optimization focuses on evaluating the dual function (\ref{dualfunction}) by solving the maximization problem with respect to $\{\boldsymbol{\Xi}_k\}_{k \in \mathcal K}$.  For fixed $\{\lambda_k,\mu_k\}_{k \in \mathcal K}$, the optimal $\boldsymbol{\Xi}_k^\star$ can be obtained in closed form via whitening transformation and generalized water-filling.

Specifically, define
\begin{align}
\mathbf B_k
&=\sigma^2\mathbf I_M+\sum_{k'\neq k}
\hat{\mathbf H}_{k'}\boldsymbol{\Xi}_{k'}\hat{\mathbf H}_{k'}^H, \\
\mathbf Z_k
&=\mathbf B_k+\hat{\mathbf H}_k\boldsymbol{\Xi}_k\hat{\mathbf H}_k^H .
\end{align}
Using $\mathrm d\log\det\mathbf Z=\mathrm{tr}(\mathbf Z^{-1}\mathrm d\mathbf Z)$,
the gradient is
\begin{equation}
\nabla_{\boldsymbol{\Xi}_k}\mathcal L
=
\frac{1}{\ln2}\hat{\mathbf H}_k^H\mathbf Z_k^{-1}\hat{\mathbf H}_k
-\mathbf A_k.
\end{equation}
The KKT condition $\nabla_{\boldsymbol{\Xi}_k}\mathcal L=0$ yields
\begin{equation}
\hat{\mathbf H}_k^H\mathbf Z_k^{-1}\hat{\mathbf H}_k
=
\mathbf A_k.
\end{equation}

Although the KKT condition above characterizes the optimal solution, it does not directly yield a closed-form expression for
$\boldsymbol{\Xi}_k$ due to the presence of $\mathbf A_k$. To decouple the matrix structure and enable an efficient solution,
we then introduce the following whitening transformation. Define the whitening transformation
\begin{equation}
\mathbf Y_k = \mathbf A_k^{1/2}\boldsymbol{\Xi}_k \mathbf A_k^{1/2},
\qquad
\boldsymbol{\Xi}_k=\mathbf A_k^{-1/2}\mathbf Y_k\mathbf A_k^{-1/2}.
\end{equation}

After applying the whitening transformation, the effective channel gram matrix becomes
\begin{equation}
\mathbf M_k
=
\mathbf A_k^{-1/2}
\hat{\mathbf H}_k^{H}\mathbf B_k^{-1}\hat{\mathbf H}_k
\mathbf A_k^{-1/2},
\end{equation}
which is Hermitian and positive semidefinite.  
Then, its eigenvalue decomposition can be 
\begin{equation}\label{eq:eig_decomp_main}
\mathbf M_k = \mathbf V_k \boldsymbol{\Theta}_k \mathbf V_k^{H},
\end{equation}
where $\boldsymbol{\Theta}_k$ is a diagonal matrix with eigenvalues $\theta_{k,i}$ on its diagonal, and $\theta_{k,i}$ denotes the eigenvalue corresponding to the $i$-th eigenvector of $\mathbf M_k $.

To take advantage of this diagonalization, we express the whitened covariance
matrix in the same eigen-basis:
\begin{equation}
\mathbf Y_k = \mathbf V_k \boldsymbol{\Delta}_k \mathbf V_k^{H},
\qquad
\boldsymbol{\Delta}_k=\mathrm{diag}(\delta_{k,i}).
\end{equation}
Substituting this representation into the Lagrangian decouples the matrix
optimization into a set of independent scalar problems:
\begin{equation}
\max_{\delta_{k,i}\ge0}
\;
\log_2(1+\theta_{k,i}\delta_{k,i}) - \delta_{k,i}.
\end{equation}
Each scalar problem admits the generalized water-filling solution
\begin{equation}\label{eq:WF_final}
\delta_{k,i}
=
\left(1-\frac{1}{\theta_{k,i}}\right)^+ .
\end{equation}
Finally, substituting $\mathbf Y_k$ back into the whitening relation yields the optimal transmit covariance:
\begin{equation}\label{eq:Xi_solution_main}
\boldsymbol{\Xi}_k^\star
=
\mathbf A_k^{-1/2}
\mathbf V_k\boldsymbol{\Delta}_k\mathbf V_k^{H}
\mathbf A_k^{-1/2}.
\end{equation}

\subsubsection{Dual Variables Optimization and Final Solution Recovery}
The optimal dual variables are then obtained by solving the dual problem
(\ref{dualproblem}). From the KKT complementary slackness conditions,
the optimal primal covariance matrices
$\{\boldsymbol{\Xi}_k^\star\}_{k \in \mathcal K}$ satisfy
\begin{align}
\lambda_k^\star
\big(\mathrm{tr}(\boldsymbol{\Xi}_k^\star)-P_{k,\max}\big)=0,\quad
\mu_k^\star
\big(\mathrm{tr}(\widetilde{\mathbf{S}}_{k,V}\boldsymbol{\Xi}_k^\star)
-Q_{\mathrm{SAR},k}\big)=0.
\end{align}
Together with the primal feasibility conditions, we have
\begin{align}
\mathrm{tr}(\boldsymbol{\Xi}_k^\star) \leq P_{k,\max},\qquad
\mathrm{tr}(\widetilde{\mathbf{S}}_{k,V}\boldsymbol{\Xi}_k^\star)
\leq Q_{\mathrm{SAR},k}.
\end{align}
Moreover, $\mathrm{tr}(\boldsymbol{\Xi}_k^\star)$ generally decreases
with $\lambda_k$, while
$\mathrm{tr}(\widetilde{\mathbf{S}}_{k,V}\boldsymbol{\Xi}_k^\star)$
decreases with $\mu_k$. Owing to this monotonic dependence, the dual
variables $\{\lambda_k,\mu_k\}_{k \in \mathcal K}$ can be efficiently
updated via a nested bisection procedure until the complementary
slackness and feasibility conditions are satisfied.

Finally, substituting the optimal dual variables
$\{\lambda_k^\star,\mu_k^\star\}$ into
\eqref{eq:Xi_solution_main} yields the optimal solution to $\mathcal{P}_A$. The overall algorithm is summarized in Algorithm~\ref{algorithm}. 

\begin{algorithm}[h] \caption{EM-aware SE Maximization Algorithm in HMIMO-assisted System}\label{end_end} \label{algorithm} \renewcommand{\algorithmicrequire}{\textbf{Input:}} \renewcommand{\algorithmicensure}{\textbf{Output:}} \begin{algorithmic}[1] \Require Green function $\mathbf{G}(\mathbf{v},\hat{\mathbf{u}}_k)$; Fourier basis $\phi_{k,\mathbf{n}}(\mathbf{R}_k^T\hat{\mathbf{u}}_k)$ and $\psi_{\mathbf{m}}(\mathbf{v})$. \Ensure System SE $R_\mathrm{sum}$. \State Initialize iteration index $\tau=0$, dual variables $\{\mu_k^{(0)}\}, \{\lambda_k^{(0)}\}, \forall k \in \mathcal K$; \Repeat \For{each user $k \in \mathcal{K}$} \State Calculate $\mathbf{V}_k^{(\tau)} $ by (\ref{eq:eig_decomp_main}); \State Obtain optimal power allocation by (\ref{eq:WF_final}); \State Update transmit covariance $\boldsymbol{\Xi}_{k}^{\tau+1}$ by (\ref{eq:Xi_solution_main}). \EndFor \State Update dual variables $\{\mu_k^{(\tau +1)}\}, \{\lambda_k^{(\tau +1)}\}$ by minimizing the dual function.
\State Set $\tau \gets \tau +1$. 
\Until{convergence of $\boldsymbol{\Xi_k}$ or maximum iterations reached.} 
\end{algorithmic}
\end{algorithm}

The proposed algorithm can be interpreted as an extension of classical water-filling to the case with SAR constraints. In contrast to conventional water-filling, SAR constraints introduce coupling through the matrix $\widetilde{\mathbf S}_{k,V}$. To address this issue, we transform the problem into an equivalent domain where the SAR constraint is absorbed into the channel structure. This leads to a modified water-filling solution, where power is allocated over an effective channel that incorporates both communication and EM exposure effects.

\section{SIMULATION RESULTS}
This section presents numerical results that evaluate the SE performance of the proposed EM-aware transmission scheme in multiuser HMIMO system, where SE is adopted as a tractable metric to reflect the fundamental performance gains. Unless otherwise stated, the simulations are conducted with $K = 4$ users operating at a carrier frequency of 10 GHz. In addition, each user is assumed to transmit $L_k$ data streams, where $L_k$ equals the number of independent spatial modes, i.e., $L_k = N_k$, $\forall k$ \cite{qian2024spectral}. The power density of the noise is $5.6 \times 10^{-6} \ \mathrm {V^2/m^2}$ \cite{9906802}. In the SAR constraint calculations, the FCC authorizes the SAR limitation of 1.6 $\mathrm{W/kg}$ averaged over 1 gram \cite{ieee2019c951}. In line with commonly adopted SAR models for microwave communications, the tissue density is assumed to be $\rho = 1000 \ \mathrm{kg/m^3}$, and the effective conductivity of biological tissue at GHz frequencies is set to $\chi_{\mathrm t} = 1 \ \mathrm{S/m}$. A region with a volume of 1 $\ \mathrm{cm^3}$ at a distance of 10$\lambda$ from the transmitting surface is randomly selected as the test volume in this paper. To evaluate the effectiveness of HMIMO and our proposed EM-aware algorithm, we choose some benchmarks, which are as follows:
\begin{itemize}
\item \textbf{Digital MIMO (DMIMO) system \cite{9906802}:} In this scheme, both the transmitting and receiving apertures are discretized into planar antenna arrays with half-wavelength spacing. The transmit current is constant over each antenna element, and the system design reduces to conventional precoding optimization under power constraints.
\item \textbf{HMIMO without SAR constraints \cite{qian2024spectral}:} This scheme considers the same HMIMO system model but ignores the SAR constraint. The transmit covariance matrices are optimized solely under the power constraint:
\begin{equation}
\max_{\{\boldsymbol{\Xi}_k\}} R_{\rm sum}(\{\boldsymbol{\Xi}_k\}), 
\quad \text{s.t. } \mathrm{tr}(\boldsymbol{\Xi}_k) \le P_{k,\max}, \ \forall k.
\end{equation}
\item \textbf{Worst case power backoff \cite{jiang2022hybrid}:} This scheme introduces a conservative scaling factor:
\begin{equation}
\alpha^{\mathrm{worst}} =
\min \left\{ 1, \frac{Q}{\mathrm{SAR}^{\mathrm{worst}}} \right\},
\end{equation}
where
\begin{equation}
\mathrm{SAR}^{\mathrm{worst}} =
\max_{k} \max_{\mathrm{tr}(\boldsymbol{\Xi}_k) \le P_{k,\max}}
 \mathrm{tr}({{\boldsymbol{\Xi}}_{k}}\widetilde{\mathbf S}_{k,V}) .
\end{equation}
The final solution is obtained by solving
\begin{equation}
\boldsymbol{\Xi}_{k,opt}^{\mathrm{wc}} =
\arg\max_{\mathrm{tr}(\boldsymbol{\Xi}_k) \le \alpha^{\mathrm{worst}}P_{k,\max}} R_{\rm sum}(\boldsymbol{\Xi}_{k}).
\end{equation}
\item \textbf{Adaptive power backoff \cite{jiang2022hybrid}:} This scheme first solves the transmit covariance optimization problem without SAR constraints:
\begin{equation}
\boldsymbol{\Xi}_k^0 = \arg\max_{\mathrm{tr}(\boldsymbol{\Xi}_k) \le P_{k,\max}} R_{\rm sum}(\boldsymbol{\Xi}_{k}), \quad \forall k.
\end{equation}
Then, the SAR constraint is enforced by scaling the transmit covariance:
\begin{equation}
\boldsymbol{\Xi}_{k,opt}^{\mathrm{adp}} = \alpha_k^{\mathrm{adp}}\boldsymbol{\Xi}_{k}^0,
\end{equation}
where the adaptive backoff factor is given by
\begin{equation}
\alpha_k^{\mathrm{adp}} =
\min \left\{ 1, \frac{Q}{\mathrm{tr}({{\boldsymbol{\Xi}}_{k}^0}\widetilde{\mathbf S}_{k,V}))} \right\}.
\end{equation}
\end{itemize}

To evaluate the computational efficiency and convergence behavior of the proposed algorithm, we first analyze its computational complexity. Let $T$ denote the number of outer iterations. In addition, let $I_\lambda$ and $I_\mu$ denote the numbers of bisection iterations required to update the dual variables $\lambda$ and $\mu$, respectively. The overall computational complexity is given by
\begin{equation}
\mathcal{O}\left(
T K
\left(
M^3
+ K (M^2 d + M d^2)
+ I_\lambda I_\mu d^3
\right)
\right).
\end{equation}
where $d = d_k = L_k N_k$ is the dimension of the transmit covariance matrix for all the users.
To provide a quantitative assessment of the computational cost, we further evaluate the runtime under a representative setting ($N_k=25$, $M=441$, $K=4$). The proposed algorithm typically converges within approximately $T \approx 7$ iterations, with a total runtime of about $2.7 \times 10^3$ seconds.

\begin{figure}[H]
	\centering
	\includegraphics[width=0.8\linewidth]{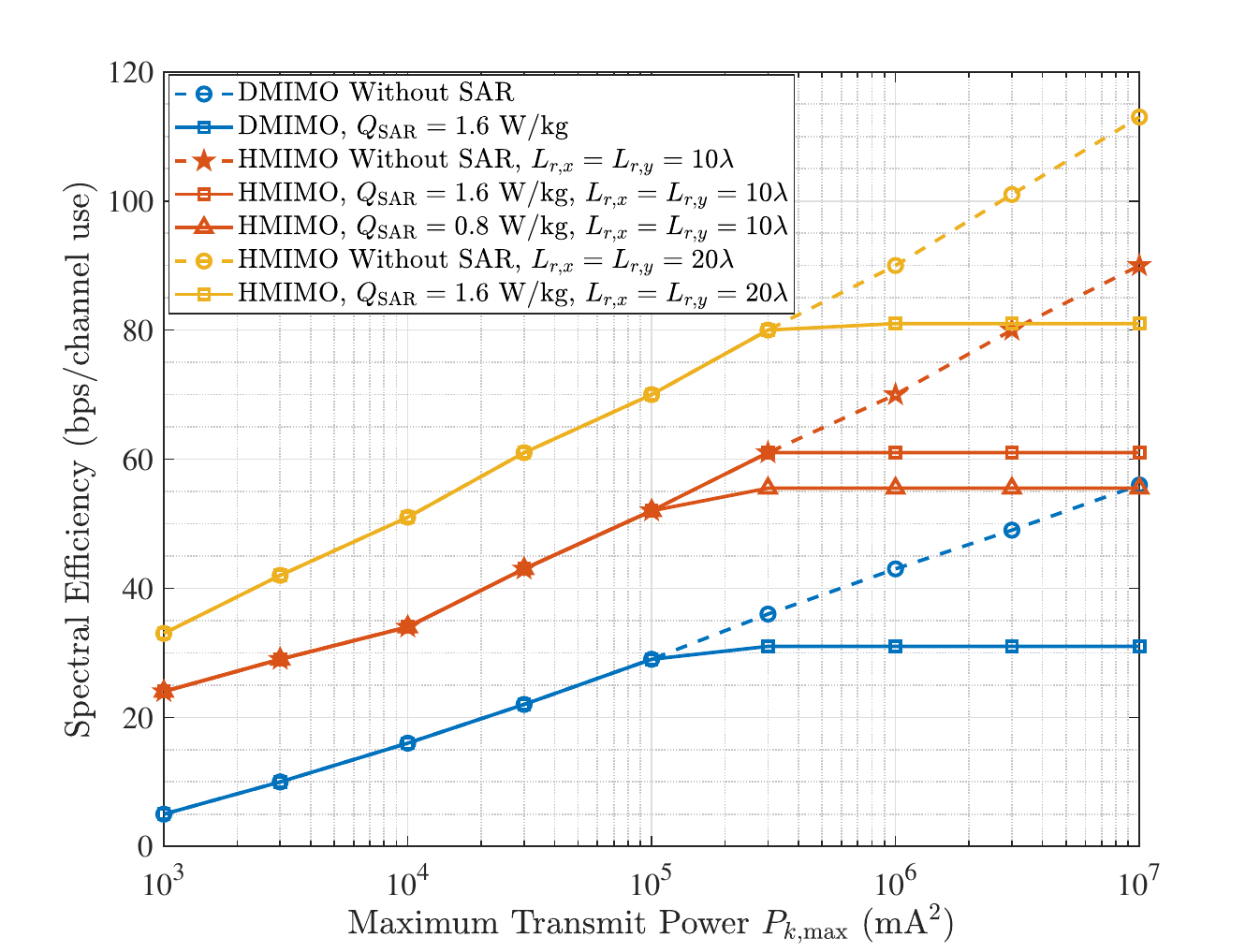}
	\caption{\ Comparison of SE versus $P_{k,\mathrm{max}}$ with different schemes.}
	\label{fig:1}
\end{figure}

To investigate the impact of SAR constraints on system performance, Fig.~\ref{fig:1} illustrates SE versus the transmit power budget $P_{k,\rm max}$, with and without SAR constraints. In the low-power regime, the SAR constraint is inactive, and the SE increases monotonically with $P_{k,\rm max}$, primarily limited by the transmit power. As $P_{k,\rm max}$ increases, the SAR constraint becomes active and gradually dominates the system behavior, leading to a saturation of SE due to EM exposure limitations. This saturation phenomenon is observed in both DMIMO and HMIMO systems, with the transition threshold determined by the SAR limit and array size. Moreover, under the same array aperture, HMIMO consistently outperforms DMIMO across the entire power range, benefiting from additional spatial degrees of freedom (DoF) enabled by the quasi-continuous aperture.

\begin{figure}[H]
	\centering
	\includegraphics[width=0.8\linewidth]{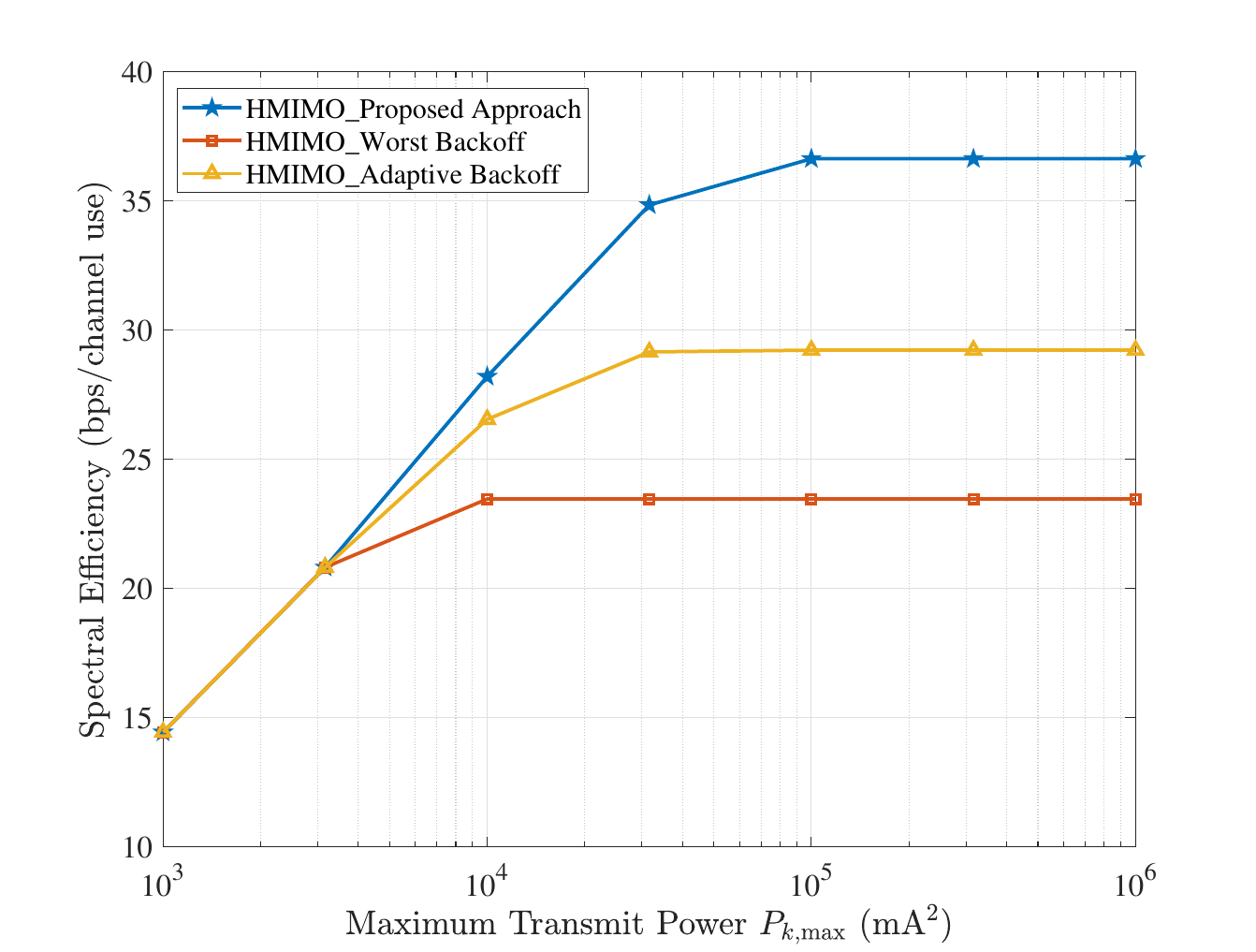}
	\caption{\ SE comparison between the proposed and baseline approaches.}
	\label{fig:2}
\end{figure}

Fig.~\ref{fig:2} further compares different SAR-aware transmission strategies. It can be observed when  $P_{k,\rm max}$ is small, all methods naturally satisfy SAR constraints. As $P_{k,\rm max}$ exceeds the threshold, SAR constraints dominate, limiting further SE improvement. Unlike baseline methods, which treat SAR as an additional power constraint, the proposed approach explicitly incorporates both power and SAR constraints, including the antenna mutual coupling effects, leading to superior SE performance.

\begin{figure}[H]
	\centering
	\includegraphics[width=0.8\linewidth]{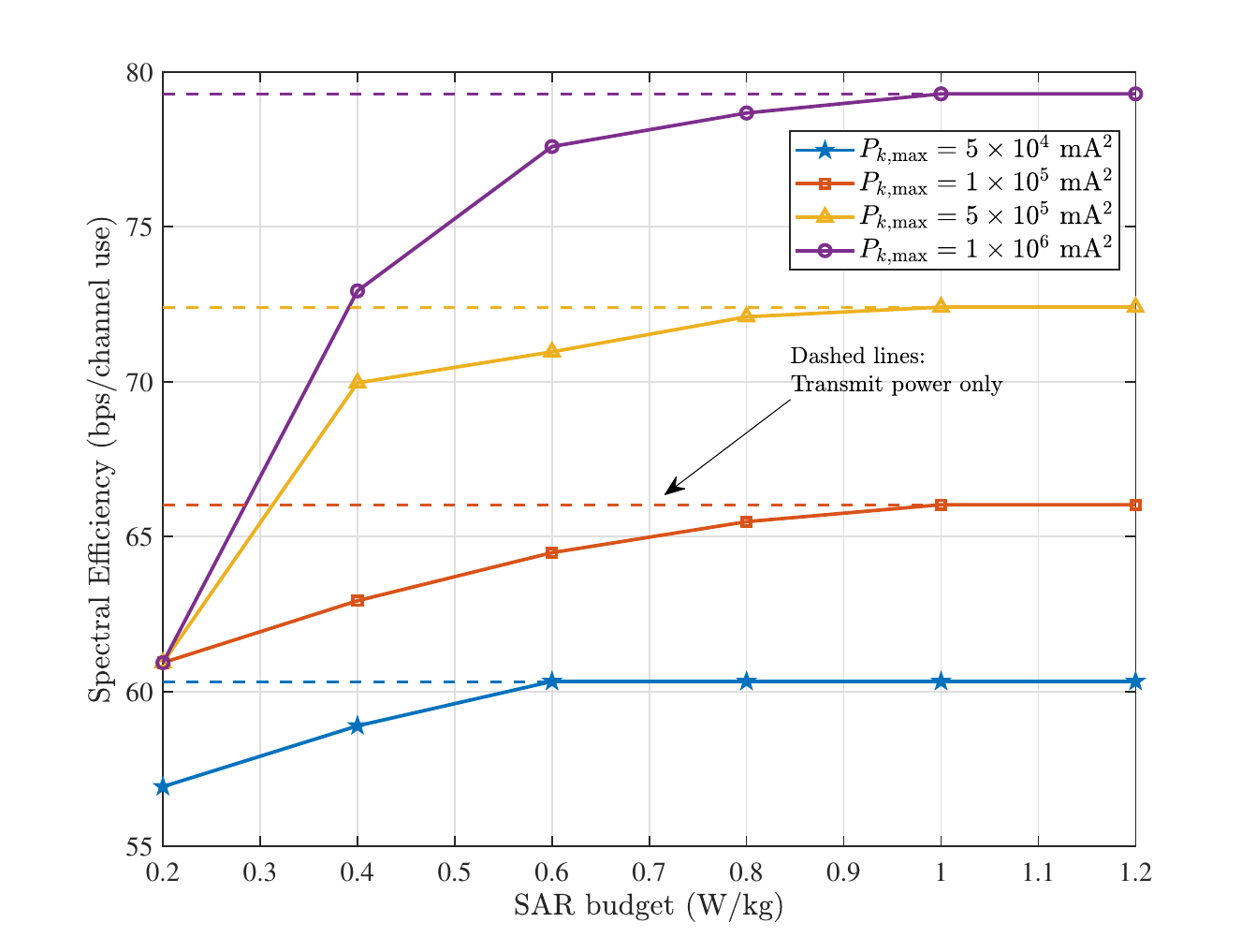}
	\caption{\ Comparison of SE versus SAR budget under different transmit power constraints.}
	\label{fig:3}
\end{figure}

Fig.~\ref{fig:3} shows the SE versus the SAR budget under different transmit power constraints. It is evident that SE increases with the SAR budget for all cases, since a relaxed EM exposure constraint allows more flexible transmit covariance design. The performance gain is more obvious in the low-SAR region, where the SAR constraint is the dominant limiting factor. As the SAR budget increases, the curves gradually approach a plateau, indicating a transition from a SAR-limited regime to a power-limited regime. The dashed lines represent the performance upper bound under power-only constraints. The convergence of all curves to these bounds confirms that the impact of SAR constraints becomes negligible at sufficiently high SAR budgets.

\begin{figure}[H]
	\centering
	\includegraphics[width=0.8\linewidth]{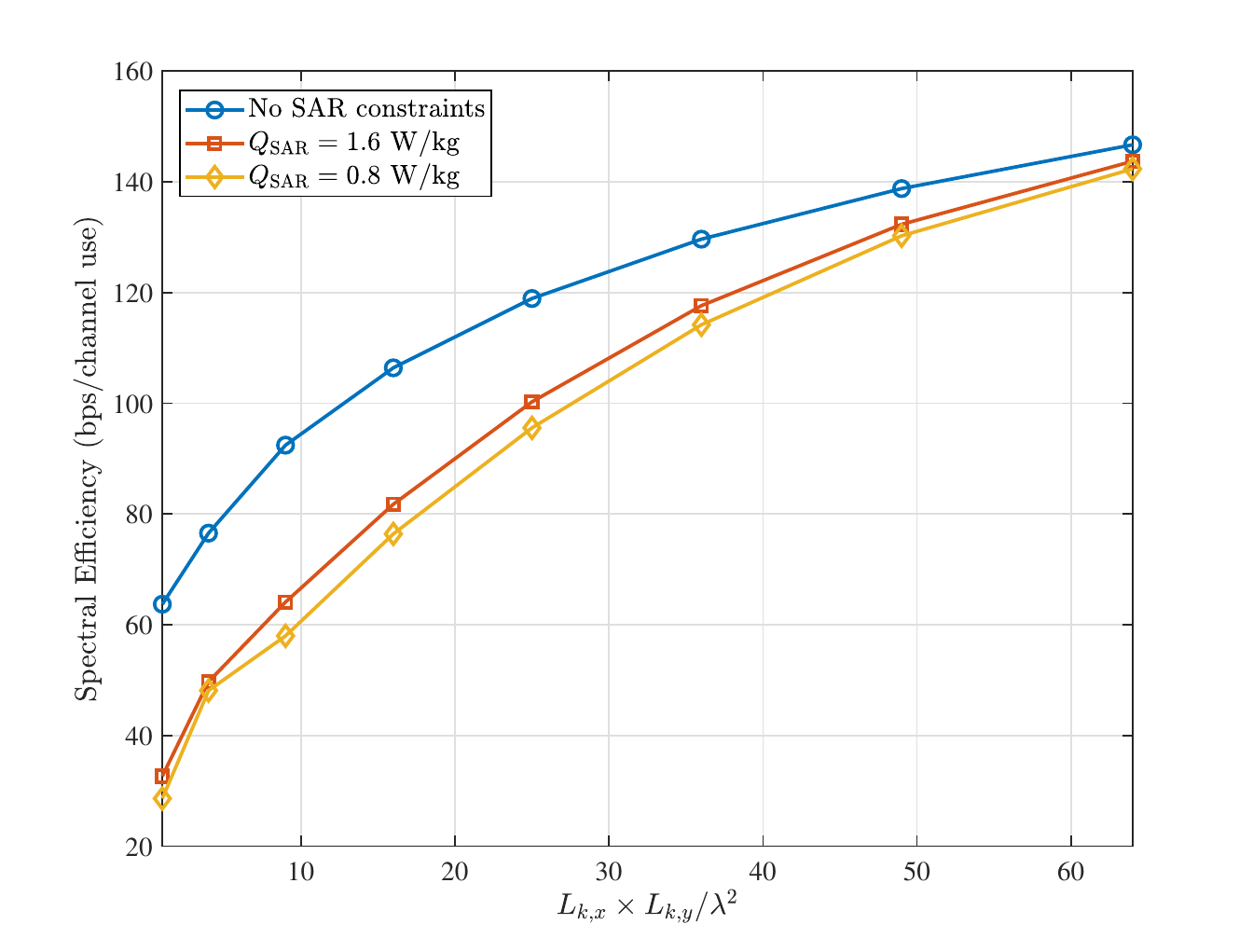}
	\caption{\ Comparison of SE versus array size under different SAR constraints.}
	\label{fig:4}
\end{figure}

Fig. \ref{fig:4} illustrates the system SE versus the array size under different SAR constraints, where the transmit power is kept the same across all schemes. It can be observed that SE increases monotonically with the array size due to enhanced spatial DoF, while gradually saturating in the large-array regime. Moreover, SAR constraints lead to a noticeable SE degradation, with stricter constraints resulting in lower performance, as they restrict the spatial distribution of transmit power and limit beamforming flexibility. However, the performance gap between different SAR levels diminishes as the array size increases, suggesting that the additional spatial DoF in large-scale HMIMO can effectively mitigate the impact of SAR constraints.

\section{Conclusion}\label{sec:Conclusion}
This work provided theoretical foundation and algorithmic support for the design of high-performance and EM-safe HMIMO systems. A modified iterative water-filling algorithm was introduced to optimize the transmit covariance matrix, subject to both power and SAR constraints. Simulation results showed that the proposed algorithm achieved fast convergence, effectively mitigated SAR-induced performance degradation compared with traditional power-backoff schemes in terms of SE. These findings highlight the practical feasibility of integrating SAR constraints into system design, offering a promising approach for future EMF-compliant HMIMO systems. 

\section*{Acknowledgment}
The authors would like to thank the editor and anonymous reviewers for their constructive comments and suggestions.

\vspace{-0.3em}

\bibliographystyle{IEEEtran.bst}
\bibliography{Refabrv_20180802,ref.bib}

\end{document}

%% file: newcommand.tex
\renewcommand{\algorithmicrequire}{\textbf{Input:}} 
\renewcommand{\algorithmicensure}{\textbf{Output:}} 